\documentclass[12pt,preprint]{aastex}

\usepackage{amsmath}
\usepackage{graphicx}
\usepackage{multirow}
\usepackage{color}
\usepackage{natbib}
\begin{document}

\title{Inclination Mixing in the Classical Kuiper Belt}

\author{Kathryn Volk and Renu Malhotra}
\affil{Lunar and Planetary Laboratory, University of Arizona, Tucson, AZ 85721, USA.}
\email{kvolk@lpl.arizona.edu}

\begin{abstract}

We investigate the long-term evolution of the inclinations of the known classical and resonant Kuiper belt objects (KBOs).  This is partially motivated by the observed bimodal inclination distribution and by the putative physical differences between the low- and high-inclination populations.  We find that some classical KBOs undergo large changes in inclination over gigayear timescales, which means that a current member of the low-inclination population may have been in the high-inclination population in the past, and vice versa.  The dynamical mechanisms responsible for the time-variability of inclinations are predominantly distant encounters with Neptune and chaotic diffusion near the boundaries of mean motion resonances.  We reassess the correlations between inclination and physical properties including inclination time-variability.  We find that the size-inclination and color-inclination correlations are less statistically significant than previously reported (mostly due to the increased size of the data set since previous works with some contribution from inclination variability).  The time-variability of inclinations does not change the previous finding that binary classical KBOs have lower inclinations than non-binary objects.  Our study of resonant objects in the classical Kuiper belt region includes objects in the 3:2, 7:4, 2:1, and eight higher-order mean motion resonances.  We find that these objects (some of which were previously classified as non-resonant) undergo larger changes in inclination compared to the non-resonant population, indicating that their current inclinations are not generally representative of their original inclinations.  They are also less stable on gigayear timescales.  
  
\end{abstract}


\section{Introduction}\label{s:Introduction}

The orbital structure of the Kuiper belt contains information about both the formation of the solar system and the Kuiper belt's subsequent evolution under the current solar system architecture.  The effect of the latter on the Kuiper belt must be well understood in order to determine how planetary formation and migration sculpted the orbital distribution of the primordial Kuiper belt (see review article by \citet{Morbidelli2008}).  Understanding the evolution of the inclination distribution of the classical Kuiper belt is of particular interest because the bimodal inclination distribution found by \citet{Brown2001} is an indication that classical Kuiper belt objects (CKBOs) might actually be comprised of two overlapping populations:  a primordial, low-inclination population, and a dynamically excited, high-inclination population (see, for example, \citet{Kavelaars2008}).  

There have been several studies attempting to identify an explanation for the bimodal inclination distribution. \citet{Kuchner2002} investigated the inclination-dependence of the long-term dynamical stability of Kuiper belt objects by performing numerical integrations of test particles with semimajor axes spread from 41 to 47 AU with uniform inclination, $i$, up to $\sim30^{\circ}$.  They found that in the inner part of the Kuiper belt, secular resonances systematically destabilize low-$i$ test particles, resulting in an inclination distribution skewed toward higher $i$; however, they did not find any mechanisms that could raise the inclinations of an initially low-$i$ population to values as large as $30^\circ$ which have been observed in the Kuiper belt.  \citet{Gomes2003} investigated how the outward migration of Neptune proposed by \citet{Malhotra1993,Malhotra1995} could scatter objects from $\sim25$ AU onto high-$i$ orbits in what is now the classical Kuiper belt region.  This study suggested that the high-$i$ population formed closer to the sun and was emplaced in the classical Kuiper belt during planetary migration, whereas the low-eccentricity, low-inclination part of the classical Kuiper belt represents a primordial, relatively undisturbed population.  It also led to the speculation that the observations of a correlation between inclinations and color in the classical Kuiper belt \citep{Jewitt2001,Trujillo2002} might be due to different origins for the objects rather than environmental effects (see review article by \citet{Doressoundiram2008}).  In addition to the color-inclination correlation, there have been correlations proposed between inclination and other physical properties such as size, albedo, and binarity \citep{Levison2001,Brucker2009,Noll2008}.  

In this paper, we investigate the degree to which planetary perturbations may have caused dynamical changes in the inclinations of both the resonant and non-resonant KBOs in the classical Kuiper belt region over gigayear timescales. If the inclination-physical property correlations are real and are primordial, then the observed low- and high-$i$ CKBOs cannot have undergone much dynamical change in inclination, else the correlations would be erased.  If the differing physical properties are due to environmental effects (such as space weathering or collisions), the effect that changes in $i$ could have on the correlations is less clear because it would depend on the timescale of the change in $i$ and how quickly the environmental effect operates to change the properties of an object.  In either case, the goal of this paper is to quantify how much mixing there is likely to be between the currently observed low- and high-$i$ populations solely as a result of dynamical evolution under the current architecture of the solar system and to quantify how this might affect the observed correlations between physical properties and inclination.  We outline our approach to modeling the current population in Section~\ref{s:modeling} and present the results of our numerical simulations in Section~\ref{s:results}.  In Section~\ref{s:discussion} we describe the dynamical mechanisms responsible for the changes in inclination for both the resonant and non-resonant populations, and we apply the results of our simulations to assess the correlations between physical properties and $i$.  Section~\ref{s:summary} gives a summary of our results.

\section{Modeling the Classical and Resonant Populations}\label{s:modeling}

To build a model of the classical Kuiper belt, we started with the observed set of KBOs with semimajor axes, $a$, in the range $33 < a < 50$ AU and that have been observed at three or more oppositions, as these objects have more accurate orbit determinations than objects with fewer observations.  Orbital elements and observational data for all objects used in this study were obtained from the Minor Planet Center (MPC) website\footnote{http://www.cfa.harvard.edu/iau/mpc.html} on 2008 May 15;  598 objects from this database met our above criteria.  We generated 10 clones of each of these objects by varying the object's nominal determined orbit by small amounts (about one part in $10^4$, comparable to the uncertainties in the orbital elements) in $a$, eccentricity, $e$, argument of perihelion, $\omega$, and mean anomaly.  

With the set of initial conditions obtained above, we performed a 10 Myr numerical integration of the orbital evolution under the current architecture of the solar system.  The integrations were performed with a mixed variable symplectic integrator based on the algorithm of \cite{Wisdom1991}.  The KBOs were treated as massless test particles moving under the gravitational field of the Sun and the four outer planets; the mass of the Sun was augmented with the total mass of the terrestrial planets.  Initial conditions for the planets were taken from the JPL Horizons service\footnote{http://ssd.jpl.nasa.gov/?horizons}~\citep{jplhorizons}.  We output the orbital elements every 5000 yr.  We then checked for stability, for membership in mean motion resonances (MMRs) with Neptune, and we calculated the average values and $rms$ variations of the orbital elements over this time period.  Any KBOs for which the majority of the clones approach within a Hill radius of any planet during this 10 Myr period are discarded from our model population because such objects will generally leave the classical Kuiper belt population on short timescales after such an encounter; objects like these with dynamical lifetimes less than 10 Myr are best classified as Centaurs \citep{Tiscareno2003} and are therefore not part of the classical Kuiper belt.  Only five of the objects were found to be unstable on Myr timescales.

\begin{figure}
   \centering
   \includegraphics{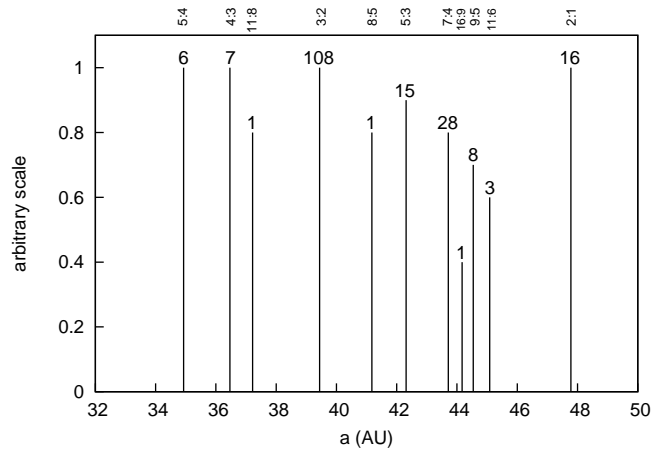} 
   \caption{Locations of identified mean motion resonances in the classical Kuiper belt.  The vertical height associated with each resonance is arbitrarily scaled to the order of the resonance.  The numbers above the lines are how many objects we have identified in each resonance.}
   \label{f:res_tree}
\end{figure}

The membership in MMRs is determined as follows.  An object is deemed resonant if its clones collectively spend more than 50\% of the 10 Myr simulation with a librating resonance angle.  While 50\% is an arbitrary choice, in practice most objects are very obviously resonant (almost all of the clones librate for the entire 10 Myr simulation).  The MMRs with Neptune (labeled as $q:p$ MMRs) are characterized by resonant angles of the form
\begin{equation}\label{eq:full_phi} 
\phi = q\lambda_{kbo} - p\lambda_{N} - r_{kbo}\varpi_{kbo}-r_{N}\varpi_{N}-s_{kbo}\Omega_{kbo}-s_{N}\Omega_{N}
\end{equation} 
where $\lambda,\varpi$, and $\Omega$ refer to the mean longitude, longitude of perihelion and longitude of ascending node, respectively, the subscripts $kbo$ and $N$ refer to the elements of a KBO and to Neptune, respectively, and $p,q,r_{kbo},r_N,s_{kbo},s_N$ are integers.  Rotational invariance imposes the constraint $q - p- r_{kbo}-r_{N}-s_{kbo}-s_{N}=0$.  For objects with semimajor axes greater than Neptune's, the MMRs of interest have $q>p>0$, and the order of the resonance is defined as $|q-p|$.   For small $e$ and $i$, perturbation theory informs that the strength of an MMR is proportional to 
$e_{kbo}^{|r_{kbo}|}e^{|r_{N}|}(\sin i_{kbo})^{|s_{kbo}|}(\sin i_{N})^{|s_{N}|}$ 
\citep{MurrayDermott1999}. Because $r_{kbo}+r_{N}+s_{kbo}+s_{N}=q-p$, the latter value is referred to as the order of a resonance, and it is a measure of the strength of the resonance.  For small or moderate values of $e$ and $\sin i$, the lower order resonances are the stronger resonances.  To simplify the process of checking for membership in the $q:p$ MMRs, we limit ourselves to resonance angles of the form
\begin{equation}\label{eq:phi}
\phi = q\lambda_{kbo} - p\lambda_{N} - (q-p)\varpi_{kbo}.
\end{equation}
Previous works \citep{Chiang2003,Elliot2005} have found no examples of KBOs in an MMR where some other allowed combination of $r_{kbo},r_N,s_{kbo},$ and $s_N$ in Equation~\ref{eq:full_phi} showed libration and the above angle did not; a spot check of different resonance angles for our simulation supports this simplifying assumption.  We identify 194 resonant objects including 108 objects in the 3:2 MMR, 28 in the 7:4 MMR, 16 in the 2:1 MMR, and 42 amongst numerous other resonances;  the highest order resonance identified is $7^{th}$ order in the KBO's eccentricity.  The locations of the MMRs identified in this population, as well as the numbers of objects in each MMR, are shown in Figure~\ref{f:res_tree}.  Table~\ref{t:res_objects} lists the designations of all the objects found in resonances, noting objects not previously identified as resonant by \citet{Gladman2008}, \citet{Lykawka2007}, or \citet{Chiang2003}.

From the remaining non-resonant objects, we wish to build an orbital distribution that represents the intrinsic distribution for the classical belt by accounting for biases in the set of observed objects.  Debiasing the observations requires knowledge of the discovery circumstances for each object, therefore in the non-resonant case we limit ourselves to objects discovered by well characterized surveys; Table~\ref{t:surveys} gives a complete list of surveys used in this work. This criterion yields 278 CKBOs.  Following \citet{Volk2008}, the debiasing is achieved in our model by assigning each object's clones a weighting factor that is determined by the probability of detection for the object;  KBOs with orbits that make them less probable to detect are given larger weights because observations thus far have a high probability of having missed many similar objects.  KBOs with orbits that keep them nearly always within the observational range of telescopic surveys are weighted less because it is likely that almost all similar objects are within our observational data set.

The primary assumption of our debiasing procedure is that an object's detectability only depends on the limiting magnitude and ecliptic latitude ranges of the observational survey.  This is a fair assumption for non-resonant objects but not for the resonant ones.  The debiasing of resonant objects is more problematic because the probability of their detection depends not only on the limiting magnitude and ecliptic latitude of the observational survey, but also on the longitude of the observations with respect to Neptune;  the libration of the resonant argument restricts the object's longitude relative to Neptune and therefore the detection probability varies with longitude \citep{Kavelaars2008}.  This is further complicated by the possibility of the Kozai resonance within the mean motion resonance (discussed further in section~\ref{sss:resonant}), which restricts where the location of the KBO's perihelion will occur relative to its longitude of ascending node, and therefore relative to the ecliptic plane \citep{Kavelaars2009}.  For example, Pluto is in both the 3:2 mean motion resonance and the Kozai resonance. The latter causes Pluto's argument of perihelion to librate around 90 degrees. Physically, this means that the location of Pluto's perihelion librates around its maximum excursion above the ecliptic plane.  This phenomenon adds to the bias against detecting such objects because they are furthest from the ecliptic plane when they are brightest.  Without knowledge of all the details of an object's resonant behavior, as well as a very specific pointing history for the observational survey, it is impossible to correctly calculate a discovery probability (see, for example,  \citet{Chiang2002} for a discussion of how discovery probabilities for the 3:2 and 2:1 MMRs depend on resonant behavior).  This makes an accurate debiased model of their intrinsic orbital distribution beyond the scope of this paper;  instead we separate the analysis of the resonant objects from the non-resonant and only apply our debiasing procedure to the non-resonant objects.  

\begin{figure}
   \centering
   \includegraphics{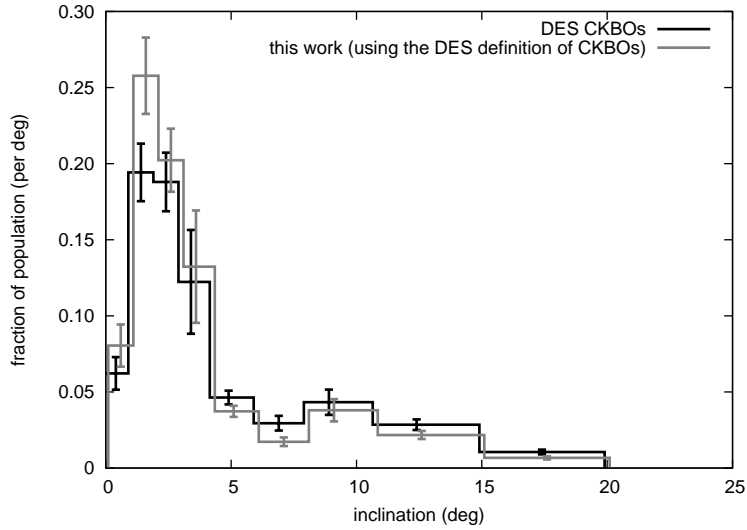} 
   \caption{Comparison of our debiasing procedure to that from the Deep Ecliptic Survey (DES) \citep{Gulbis2010}.  The distributions contain 150 DES KBOs that meet the DES definition of CKBO and are present in the data sets for this work as well as that of \citet{Gulbis2010}.  The dark line shows the debiased $i$-distribution for these KBOs from \citet{Gulbis2010} and the grey line shows our debiased distribution for the same objects.  The debiased $i$-distribution resulting from the definition of CKBOs used in this work is shown in Figure~\ref{f:itime}.}

   \label{f:compare_idist}
\end{figure}

To test our debiasing procedure, we compare our debiased inclination distribution to the debiased inclination distribution reported for CKBOs from the Deep Ecliptic Survey (DES) by \citet{Gulbis2010}.  Because the DES classification scheme differs significantly from our own, we only compare the subset of our CKBOs that overlaps with their definition.  Under the DES classification scheme \citep{Elliot2005}, CKBOs are non-resonant KBOs with eccentricities $< 0.2$ and Tisserand parameters with respect to Neptune $>3$.  Using this definition, \citet{Gulbis2010} include 150 objects in their debiased CKBO $i$ distribution that are also included within our CKBO dataset;  the debiased inclination distribution for these 150 CKBOs is shown in Figure~\ref{f:compare_idist}.  Following \citet{Gulbis2010}, we smooth our debiased $i$-distribution at higher inclinations (where there are fewer observations on which to base the debiased distribution) by using variable bin sizes:  $1^{\circ}$ for $i \le 4^{\circ}$, $2^{\circ}$ for $4 < i \le 10^{\circ}$, and $5^{\circ}$ for $i >10^{\circ}$.  The entire distribution is normalized to 1 and reported in Figure~\ref{f:compare_idist} as the fraction of CKBOs per degree in inclination.  We performed a Kolmogorov-Smirnov (K-S) test to quantify the similarity of the two debiased inclination distributions.  The K-S test measures the maximum difference, D, between the cumulative distributions of two individual data sets and then evaluates the probability that the two are drawn from the same parent distribution \citep{Press1992}.  The K-S test comparing the debiased CKBO $i$ distribution from \citet{Gulbis2010} to our debiased $i$ distribution for the 150 CKBOs in our sample that overlap with the DES sample yields a 65\% probability that the two distributions are the same.  Compared to the DES debiased distribution, our distribution is slightly systematically skewed toward lower inclinations.  This difference is due to our simplifying assumptions in the debiasing procedure: whereas \citet{Gulbis2010} use the complete details of each discovery image (such as limiting magnitude and ecliptic latitude) to calculate debiasing factors, we used the DES survey averages described by \citet{Elliot2005}.  Given this limitation, our debaised distribution is close enough to that of \citet{Gulbis2010} to give us confidence in our debiasing procedure. A further test of our debiasing procedure is to compare the resulting inclination distribution to that found for the classical Kuiper belt by \citet{Brown2001}, who used a definition of CKBOs similar to our own;  we discuss this comparison in Section~\ref{s:results}.

The set of 194 resonant and 278 non-resonant objects in the classical belt region selected from the 10 My simulation were then integrated for 4 Gyr with output every 1 Myr to study the long-term orbital evolution of each object under the perturbations of the four outer planets. As in the case of the 10 Myr integrations, 10 clones of each object were integrated as massless test particles, and we terminated the integration of any particle that approached within one Hill sphere of any planet during the 4 Gyr simulation.  In the analysis in the following section, we refer to the particles that survive to the end of the 4 Gyr simulation as ``stable", and the others (that have dynamical lifetimes less than 4 Gyr) as ``unstable".  The vast majority of the unstable particles in our sample have gigayear lifetimes, and are thus expected to be representative of the presently existing Kuiper belt; we include their contributions in our time-averaged results.

\section{Analysis and Results}\label{s:results}

\begin{figure}
  \centering
   \includegraphics{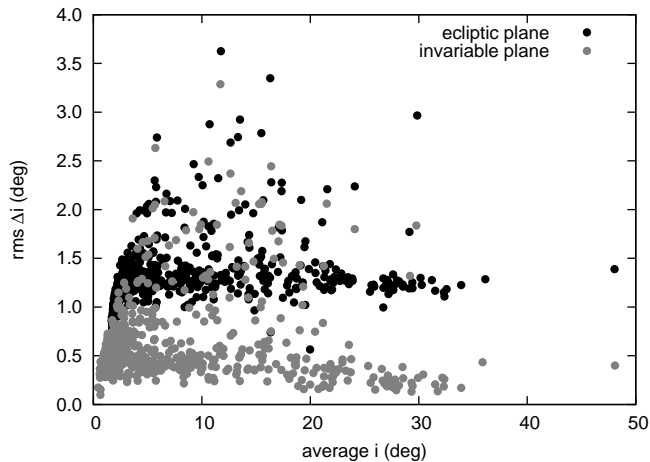} 
   \caption{Variation in inclination vs. average inclination for the initial 10 Myr integration relative to the ecliptic plane (black) and the invariable reference plane (gray).}
   \label{f:i_di}
\end{figure}

To analyze the inclination evolution of our test particles, we must choose an appropriate coordinate system for our results.  The standard output for the numerical integrations are heliocentric orbital elements; in our case these are referenced to the J2000 coordinate system, meaning all inclinations are measured relative to the ecliptic plane of the epoch.  Because we are interested in changes in the inclinations of the test particles, it is preferable to measure those inclinations relative to a more physically meaningful reference plane;  the best reference plane for determining changes in the orbits of the KBOs would be the mean plane of the Kuiper belt.  Several works have calculated the mean plane of the Kuiper belt based on observations and secular perturbation theory \citep{Elliot2005,Brown2004,Chiang2008}.  In general, the average plane  varies with semimajor axis \citep{Chiang2008}; this is not a desirable feature for a reference plane for our numerical simulations.  A more convenient and easily calculated reference plane is the invariable plane of the solar system.  Both \citet{Chiang2008} and \citet{Elliot2005} find only small deviations between the invariable plane and the average Kuiper belt plane;  \citet{Brown2004} found a larger difference between these planes, but the difference was still less than $1^{\circ}$.  We will therefore reference all KBO inclinations in this paper to the invariable plane.  To do this, we calculate the location of the barycenter for the Sun and the four outer planets, and then we calculate the total angular momentum vector for Sun and outer planets relative to the barycenter; the plane normal to this vector becomes our reference plane for the orbital inclinations.  Adopting the invariable plane diminishes the amplitude of the short-term oscillations in the KBO test particles' inclinations compared to when the ecliptic plane is used.  This is because their short-term inclination evolution is driven by the short-term inclination evolution of the planets.  If the ecliptic plane is used as the reference plane, the planets' inclinations average $1.5-2^{\circ}$ over Myr timescales;  if the invariable plane is used instead, this average is $0.3-1^{\circ}$.  The reduction in the amplitude of the corresponding changes in the test particles' inclinations is illustrated in Figure~\ref{f:i_di}, which plots the $rms$ variation of each test particle's inclination vs. the mean inclination for the 10 Myr simulation.  This reduction in the `typical' spread in the inclinations expected over Myr timescales makes it easier to distinguish the test particles with larger inclination changes in the longer simulation.  Unless noted otherwise, all inclinations in this paper are referenced to the invariable plane.

\begin{figure}
   \centering
   \includegraphics{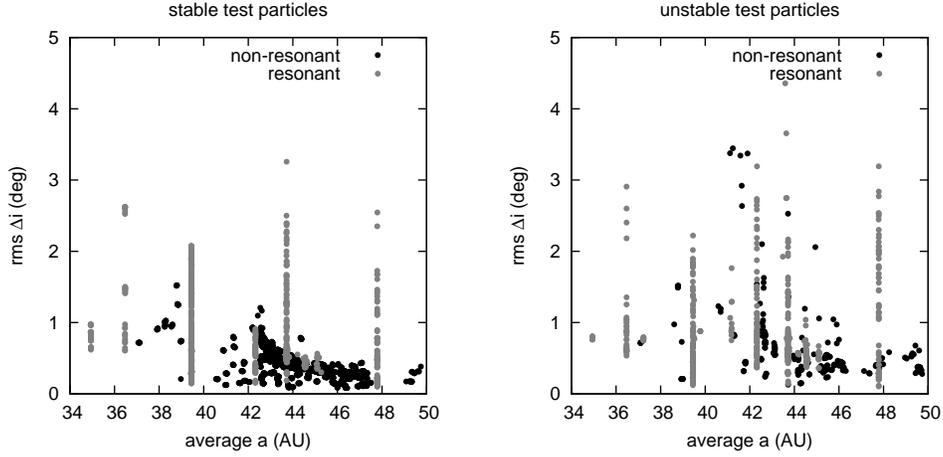} 
   \caption{Variation in inclination as a function of average semimajor axis for the non-resonant (black) and resonant (gray) test particles over the 10 Myr integration.  The left panel shows test particles that survive the 4 Gyr simulation without a close encounter with Neptune, and the right panel shows those that are removed from the simulation by a close encounter.}
   \label{f:a_10Myrdi}
\end{figure}

We can use several measures to determine the mobility of the KBOs in inclination space.  Figures~\ref{f:a_10Myrdi} and~\ref{f:a_di} show the $rms$ variation in inclination over the initial 10 Myr simulation and the 4 Gyr simulation, respectively, for all the test particles as a function of average semimajor axis over the same time period (note the different scales on the y-axes of Figures~\ref{f:a_10Myrdi} and~\ref{f:a_di}).  The stable test particles (those that survive the entire 4 Gyr simulation) are shown separately from the particles that have close encounters with Neptune.  The inclination variations in the 4 Gyr simulation are generally larger than in the 10 Myr simulation, especially for the resonant objects.  The maximum changes in any one particle's inclination over the entire simulation can be many times these $rms$ values.  Figure~\ref{f:maxdi} shows the maximum $\Delta i$ for objects that survive the entire simulation and those that do not, where both groups are separated into CKBOs and resonant objects.  Not surprisingly, the unstable objects undergo the largest changes in $i$, but several of the stable CKBOs (less than 1\%) and quite a few of the stable resonant objects (8\%) show changes of more than $10^{\circ}$.  Of the total debiased CKBO population, 10\% of objects deviate from their initial 10 Myr average $i$ by more than $5^{\circ}$ at some point during the simulation (2\% of the stable CKBOs and 84\% of the unstable CKBOs), and 4\% deviate by more than $10^{\circ}$ ($<$1\% of the stable CKBOs and 40\% of the unstable CKBOs).  For the observed population of resonant objects, these percentages are 51\% (38\% of the stable objects and 87\% of the unstable objects) and 23\% (8\% of the stable objects and 64\% of the unstable objects), respectively.

Given that some objects' inclinations are quite changeable over time due to planetary perturbations, it is important to examine the stability of the overall inclination distribution.  Figure~\ref{f:itime} shows snapshots of the inclination distribution for the debiased non-resonant CKBOs at the beginning of the simulation, 500 Myr into the simulation, and at 4 Gyr.  (Note that we use the same binning in this figure as in Fig.~2.)  The results of a K-S test (described in Section~\ref{s:modeling}) show that there is a $52\%$ probability that the initial and final snapshots are drawn from the same inclination distribution, and there is a greater than $90\%$ probability of similarity between all snapshots taken after 500 Myr.  The lower K-S probability for the early snapshots compared with the final snapshot is most likely due to the circumstance that if an object on an unstable orbit is heavily weighted in the distribution and is then either removed from the simulation or evolves onto a more stable orbit, the resulting change in the overall distribution can be statistically large.  The difference between the initial and final distribution is not large, and a K-S probability of $~\sim50\%$ is not so small as to rule out the possibility that the distributions are the same;  the 90\% K-S probability after the first 500 Myr shows that, even if some objects undergo large changes in inclination, the overall distribution is quite stationary.

The debiased inclination distribution we find for the CKBOs (Figure~\ref{f:itime}) shows a distinct peak at low-inclinations ($i < 5^{\circ}$) with a substantial part of the population spread over a wide range of higher inclinations.  The reader will note that this is significantly different from the $i$-distribution for the DES CKBOs by \citet{Gulbis2010} (see Figure~\ref{f:compare_idist}): the notable paucity of high-$i$ CKBOs in Figure~\ref{f:compare_idist} compared to Figure~\ref{f:itime} is owed entirely to the differing definitions for CKBOs.  The cutoff in the Tisserand parameter with respect to Neptune ($T > 3$) in the DES definition is the major contributor to the difference;  for example, a KBO with semimajor axis $a=40$ AU and an eccentricity of 0.1 must have an inclination less than $\sim 12^{\circ}$ and a KBO with $a=45$ AU and an eccentricity of 0.1 must have an inclination less than $\sim 16^{\circ}$, to qualify as a CKBO using the DES definition.  The eccentricity cutoff ($e<0.2$) adopted by DES also excludes a few of our high-$i$ CKBOS. \citet{Brown2001} use a CKBO definition much more similar to ours than that of the DES, the only differences being the adoption of a slightly narrower range in semimajor axes ($40 < a < 48$ AU) and the exclusion of a few of the higher eccentricity KBOs in that semimajor axis range for having perihelion distances very similar to that of scattered disk objects.  Following \citet{Brown2001}, we model our initial inclination distribution to a sum of two Gaussians multiplied by $\sin i$
\begin{equation}\label{eq:idist}
f(i) = \sin i \left[ A \exp \left( \frac{-i^2}{2 \sigma_1^2} \right) + (1-A) \exp \left( \frac{-i^2}{2 \sigma_2^2} \right) \right].
\end{equation}
 We find the following best-fit parameters:  $A = 0.95 \pm 0.02$, $\sigma_1 = 1.4 \pm 0.3^{\circ}$, and $\sigma_2 = 15 \pm 3^{\circ}$.  These may be compared with $A = 0.93 \pm 0.03$, $\sigma_1 = 2.2^{+ 0.2}_{- 0.6}$, and $\sigma_2 = 17 \pm 3^{\circ}$ from Table 1 of \citet{Brown2001}.  The largest difference between the two best-fits is in the width of the narrow component; this difference is mainly owed to the differing reference planes;  \citet{Brown2001} used the ecliptic plane as the reference plane for the inclinations, while we use the invariable plane.  Our fit indicates that 83\% of the CKBOs are in the wide Gaussian and 17\% in the narrow Gaussian, similar to the 81\% and 19\% found by \citet{Brown2001}.  The inclination at which the narrow and wide components have equal numbers of objects is $i=5^\circ$ in our model and $i=7^\circ$ in \cite{Brown2001}.  Both fits are plotted as the smooth curves in Figure~\ref{f:itime}.  While the data appear to support the existence of both a low-$i$ and a high-$i$ component in the inclination distribution, Figure~\ref{f:itime} shows that the model curves do not fit the data very well.  This poor fit is at least partially due to the noisy nature of the high-$i$ portion of our debiased distribution: there are still relatively few observed objects at large-$i$.  The poor fit could also be an indication that a double Gaussian is not a very good model for the intrinsic distribution.  \citet{Gulbis2010} try various other functional forms for fitting the debiased DES inclination distribution and find that a Gaussian plus a generalized Lorentzian provides an improved fit;  however, the improvement over the double Gaussian is small, so for ease of comparison with the \citet{Brown2001} results we only report a fit to the double Gaussian form of the inclination distribution.

From the 4 Gyr simulation we want to quantify how well an object's current inclination correlates with its past or future inclination; i.e. what is the probability that a currently `low' inclination object has always been in the `low' inclination group?  The answer to this question may depend on how `low' inclination is defined.  We adopted a cutoff inclination value, and then examined how the degree of mixing between the high- and low-inclination populations depends on the value of the cutoff inclination.  Figure~\ref{f:imixing} shows the time-averaged fraction of all non-resonant and resonant CKBOs that can be found at inclinations that are inconsistent with their initial (i.e. current epoch, 10 Myr average inclination) classification into either the high- or low-$i$ populations as a function of the $i$ cutoff that defines the low- and high-inclination populations.  For example, for a cutoff of $5^{\circ}$ (which is sometimes used to distinguish the so-called `hot' and `cold' CKBOs \citep{Doressoundiram2002,Brucker2009}) one would expect $1.5\%$ of all non-resonant CKBOs and $11\%$ of resonant objects to currently be at inclinations inconsistent with their primordial inclinations, because the populations as a whole spend $1.5\%$ and $11\%$ of their time at inclinations inconsistent with their current classifications\footnote{If, instead of using the average inclination from the 10 Myr simulation, we use the instantaneous inclination with respect to the invariable plane, the percentage of misclassified objects using a $5^{\circ}$ cutoff between low- and high-$i$ populations is still $1.5\%$ for the CKBOs, but for the resonant objects the percentage increases to 13\%.}.  Our results in Figure~\ref{f:imixing} also show that the inclination of $\sim5^{\circ}$ is justified as a divide between the hot and cold population as it corresponds to a local minimum in the `misclassification' distribution.  

We note that if the ecliptic plane were used for this calculation instead of the invariable plane, the result is more inclination mixing (and therefore a fuzzier distinction) between the two classes, as shown in the second panel of Figure~\ref{f:imixing}.  For both the CKBOs and the resonant KBOs, the percentage of objects we expect to find at inclinations inconsistent with their initial classification is smaller than the percentage of objects found to have deviated from their initial inclinations by large amounts at some point in our simulation.  This is because objects that experience large deviations in $i$ do not spend a large fraction time at the extreme values.  So even though half of the resonant KBOs will at some point have an excursion more than $5^{\circ}$ away from their initial $i$, only 12\% of these objects spend more than 10\% of their time more than $5^{\circ}$ away from their initial $i$.  Similarly, only 6\% of CKBOs spend more than 10\% of their time at values of $i$ more than $5^{\circ}$ away from their initial $i$.

\section{Discussion}\label{s:discussion}

\subsection{What drives the changes in inclination?}\label{ss:deltai}

\subsubsection{CKBOs (non-resonant)}\label{sss:ckbos}

Approximately 10\% of the total debiased CKBO population experiences a change in $i$ of more than $5^{\circ}$ during our 4 Gyr simulation.  Here we examine the causes of these relatively large changes in inclination to better understand how the inclinations of CKBOs evolve over time.  To do this we examined the time histories of the test particles exhibiting the largest overall changes in inclination as well as those exhibiting small changes.  Objects with stable orbits and inclinations tend to undergo variations in $i$ of $\sim1^{\circ}$ that occur on timescales comparable to or shorter than our output resolution of 1 Myr (an example of this is shown in Figure~\ref{f:low_i}).  This is consistent with what is expected from the secular perturbations of the planets on a massless test particle;  near Neptune, the plane of a KBO's orbit will wobble about the invariable plane with an amplitude similar to Neptune's proper inclination; the latter is $\sim0.7^{\circ}$ inclined to the invariable plane \citep{Brown2004,Chiang2008}.

\begin{figure}
   \centering
   \includegraphics{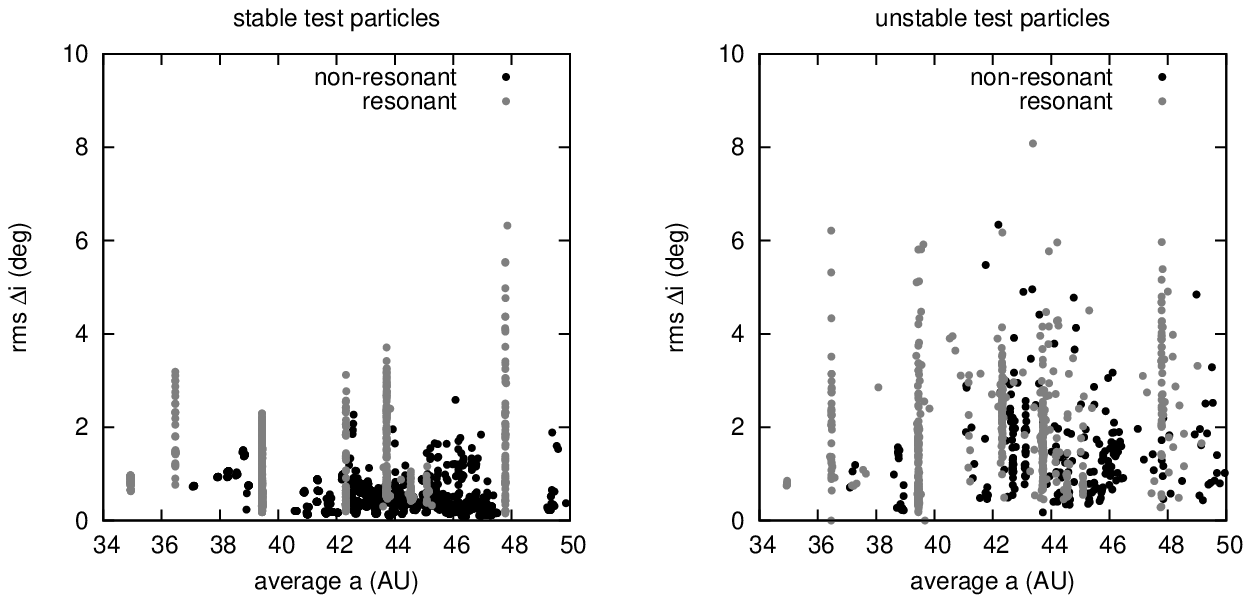} 
   \caption{Variation in inclination as a function of average semimajor axis for the non-resonant (black) and resonant (gray) test particles over the 4Gyr integration.  The left panel shows test particles that survive the 4 Gyr simulation without a close encounter with Neptune, and the right panel shows those that are removed from the simulation by a close encounter.  The 50 Myr time period prior to particle removal is excluded from the calculation of the average and $rms$ variation.}
   \label{f:a_di}
\end{figure}

\begin{figure}
   \centering
   \includegraphics{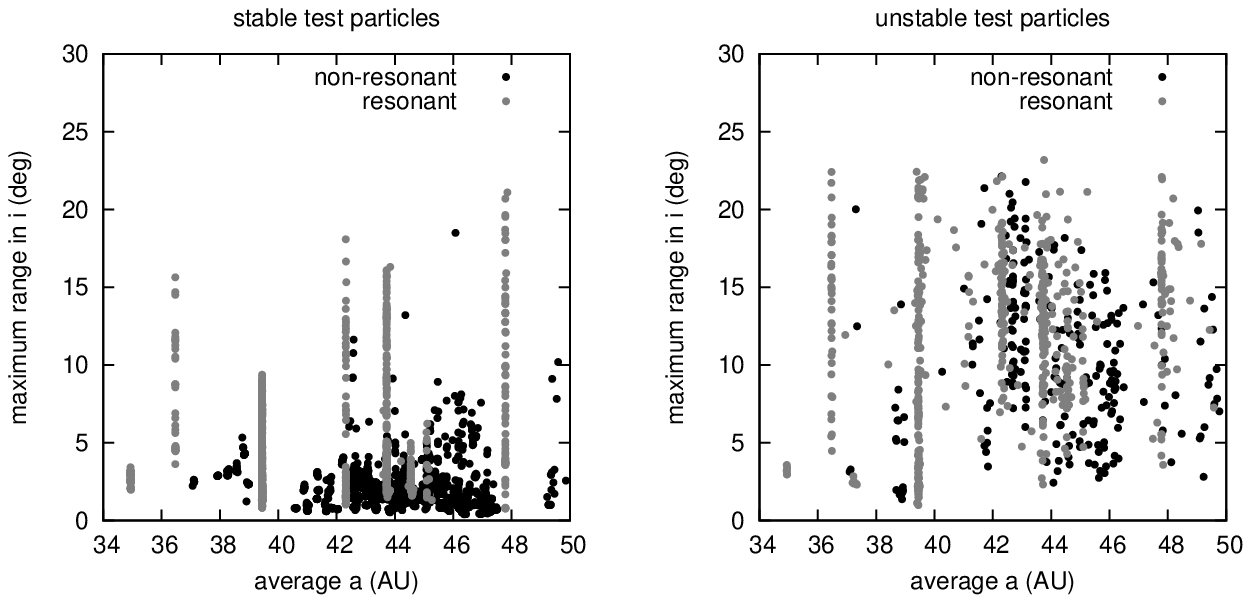} 
   \caption {Maximum range in inclination as a function of average semimajor axis for the non-resonant (black) and resonant (gray) test particles over the 4 Gyr integration.    The left panel shows test particles that survive the 4 Gyr simulation without a close encounter with Neptune, and the right panel shows those that are removed from the simulation after a close encounter.  The 50 Myr time period prior to particle removal is excluded from the calculation of the average and maximum range.}
   \label{f:maxdi}
\end{figure}

\begin{figure}
   \centering
   \includegraphics{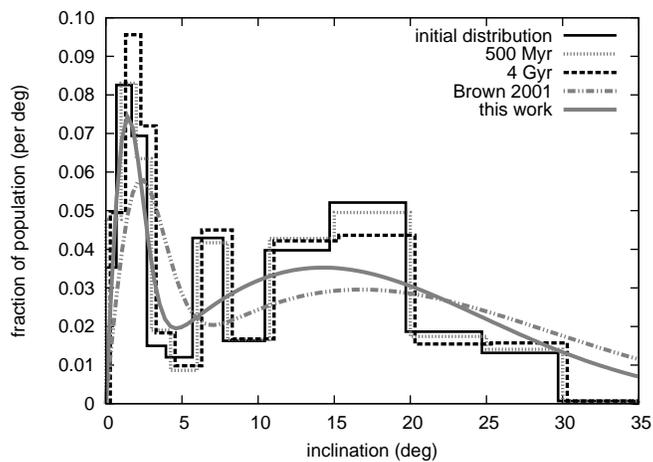} 
   \caption{Snapshots of the debiased inclination distribution for the non-resonant CKBOs taken at the beginning of the simulation, at 500 Myr, and at 4 Gyr. (Note that the $i$-distribution depends heavily on the definition used for CKBOs.  The differences between this figure and Figure~\ref{f:compare_idist} are entirely owed to the differences between our classification scheme and that used by \citep{Gulbis2010}.) The two smooth curves represent model fits of the CKBO inclination distribution to the sum of two Gaussians multiplied by $\sin i$ (Equation~\ref{eq:idist}).  The solid grey line is our best-fit model to the initial distribution, and the dashed grey line is the model of \citet{Brown2001} for comparison. }
   \label{f:itime}
\end{figure}

\begin{figure}
   \centering
   \includegraphics{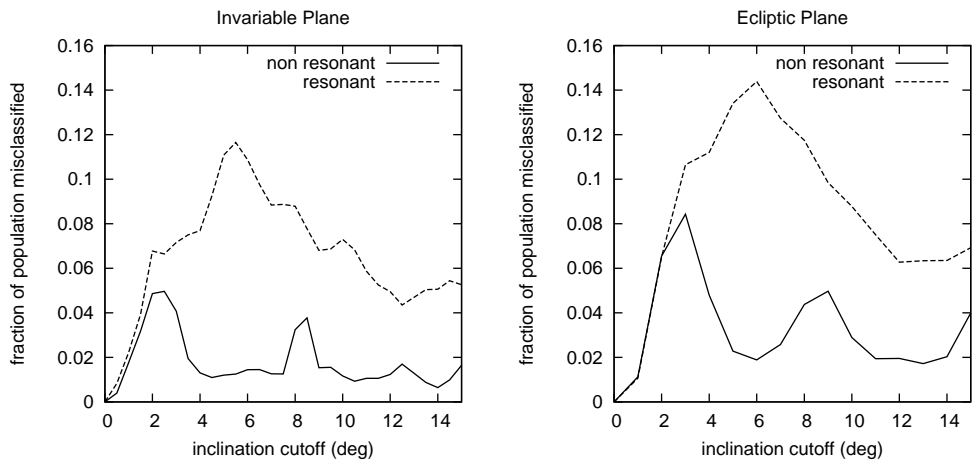} 
   \caption{Time-averaged fraction of the total observed resonant population (dashed line) and of the total debiased non-resonant population (solid line) that is misclassified.  This fraction is plotted as a function of the cutoff between the low- and high-inclination populations.  Left panel: inclinations referred to the invariable plane; Right panel: inclinations referred to the ecliptic plane. }
   \label{f:imixing}
\end{figure}

\begin{figure}
   \centering
   \includegraphics[width=5.5in]{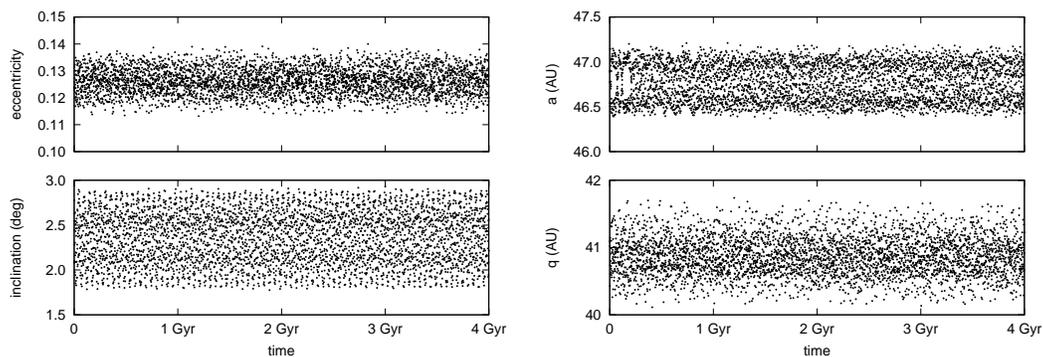} 
   \caption{Eccentricity, inclination, semimajor axis, and perihelion distance evolution over 4 Gyr for a test particle that maintains a low $i$ throughout the entire simulation.}
   \label{f:low_i}
\end{figure}

\begin{figure}
   \centering
   \includegraphics[width=5.5in]{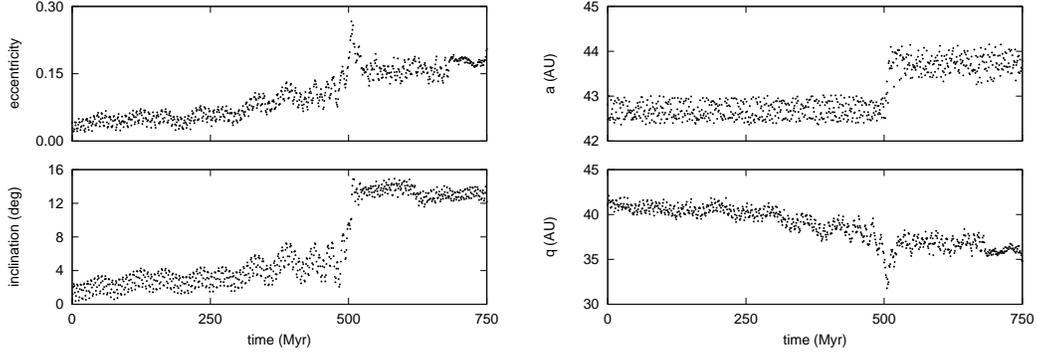} 
   \caption{Eccentricity, inclination, semimajor axis, and perihelion distance evolution over the first 750 Myr of the 4 Gyr simulation for a test particle that has a distant encounter with Neptune (at about 500 Myr) and a consequent jump in $a$, $e$, and $i$.}
   \label{f:jump}
\end{figure}

Most (about 80\%) of the classical belt test particles with large $\Delta i$ have changes in $i$ that can be attributed to distant encounters with Neptune.  The time resolution of our 4 Gyr simulation output is not fine enough to detect the encounters, but we can diagnose such encounters with Neptune because they manifest as discrete jumps in the semimajor axis, eccentricity, and inclination of the test particles, and these occur when the particle is at a local minimum in the time history of the perihelion distance.  An example is shown in Figure~\ref{f:jump}, where an $\sim8^{\circ}$ increase in $i$ occurs at a perihelion distance $q$ of about 32 AU.  These moderately distant encounters with Neptune certainly explain the large changes in $i$ for the unstable subset of CKBOs.  The values for the typical spread in $i$ and the maximum $\Delta i$ reported here exclude the 50 Myr period prior to the closest approach with Neptune (within one Hill radius) that removes a test particle from our simulation, so the variations in $i$ shown in Figures~\ref{f:a_di} and ~\ref{f:maxdi} represent changes that are occurring due to more distant encounters.  Unstable test particles can wander substantially in inclination over timescales much longer than 50 Myr before having an encounter close enough to completely remove the particle from the classical belt region.  Approximately one-third of the test particles experiencing large $\Delta i$ as a result of a distant encounter with Neptune persist in the classical belt for timescales longer than 250 Myr after the encounter.  A small subset of the stable CKBOs will also experience distant encounters with Neptune that lead to significant changes in $i$ without leading to a close enough encounter to completely remove the test particle (at least not within the 4 Gyr of our simulation).  

Approximately 20\% of the test particles (both stable and unstable) with large maximum $\Delta i$ values in the classical belt owe their inclination variability to brief stints of libration within an MMR.  Careful examination of Figures~\ref{f:a_di} and~\ref{f:maxdi} shows that some of the CKBO test particles in close proximity to resonant particles (as measured by their average semimajor axis values) also have large $rms$ variations in $i$ and/or maximum $\Delta i$, similar in magnitude to the nearby resonant particles. Some test particles are captured for long enough in MMRs that we detect libration of the resonance angle in our simulation output, and we have identified temporary capture in most of the occupied resonances and a few very high-order MMRs with Neptune such as the 19:10 and 21:11. Short-lived resonance libration is difficult to ascertain for test particles that spend less than a few tens of millions of years in resonance, due to the lower time resolution of our simulation output.   Temporary capture into various resonances in the classical belt region was also noted by \citet{Lykawka2006} in a long-term simulation of test particles started near the 7:4 MMR.  In our simulation, we find that some test particles start in a resonance, but exit the resonance without being removed from the classical belt population.  Some of the resonant test particles also experience only intermittent libration (see Section~\ref{sss:resonant}), so it is possible that objects we classify as non-resonant are close enough to a resonance boundary to experience occasional librations.  Such chaotic diffusion across resonance boundaries is the cause of many of the cases of large inclination variability of CKBOs.

A few percent of the stable particles with large $\Delta i$ in the 4 Gyr simulation cannot be explained with either of the above two mechanisms -- distant Neptune encounters or temporary resonance capture.  There is a secular resonance at $\sim41$ AU that has been shown to raise the inclinations of initially low-$i$ test particles \citep{Knezevic1991}, and, as discussed above, there are also many currently occupied MMRs in the 40 to 45 AU range which might be expected to influence the orbital evolution of nearby test particles.  Because these dynamical mechanisms work mainly in the inner part of the classical belt region ($a < 45$ AU), we expect the amplitude of changes in $i$ to decrease in the outer region, with the exception of test particles near the 2:1 MMR.  In the 10 Myr simulation, this is indeed what we find; the typical spread of $i$ for the stable objects (as seen in the left panel of Figure~\ref{f:a_10Myrdi}) is small in the outer region, and it increases moving inward toward both the 5:3 MMR at $\sim42.3$ AU and the secular resonances at 41--42 AU.  However, in the longer simulation (Figs.~\ref{f:a_di} and \ref{f:maxdi}) we see some particles with surprisingly large $rms$ variations in $i$ and maximum changes of $5-10^{\circ}$ in the semimajor axis range of 45 to 47 AU where there are no strong MMRs with Neptune and no known secular resonances.  This zone does have several higher order mean motion resonances, such as Neptune's 11:6 and 7:19, as well as a three-body resonance involving both Neptune and Uranus~\citep{Nesvorny2001}. We looked for librations of candidate resonance angles; our output cadence of 1 Myr (for the 4 Gyr simulation) can only detect libration behavior on timescales greater than $\sim10$ Myr; we did not detect any such long-lived librations.  For our test particles between 45 and 47 AU, the effects of chaos arising from the overlap of these high-order interactions may be the cause of the larger variability in $i$ that we find in the 4 Gyr simulation (but do not find in the 10 Myr simulation). This conjecture is supported by \citet{Nesvorny2001}'s finding of a small positive Lyapunov exponent in this range of semimajor axis, although their analysis was confined to very low inclination, low eccentricity orbits.  A deeper analysis of the dynamics of objects in this zone is left for a future investigation.

\begin{figure}
   \centering
   \includegraphics[width=5.5in]{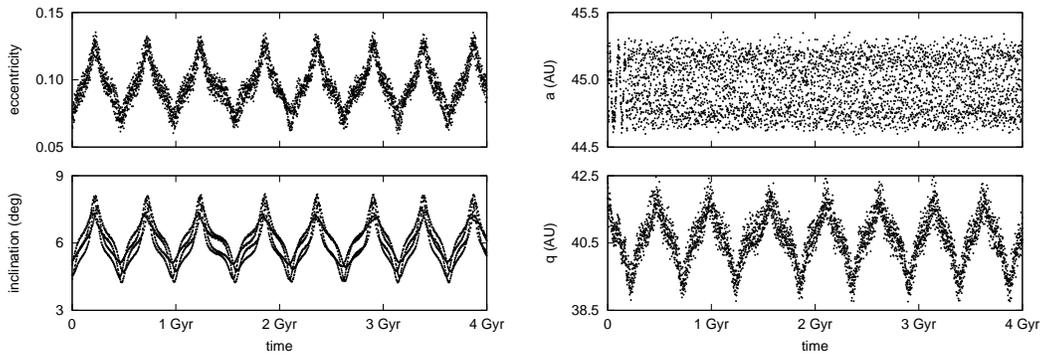} 
   \caption{Eccentricity, inclination, semimajor axis, and perihelion distance evolution over 4 Gyr for a test particle with an anomalously long-period variation in $i$ and $e$.}
   \label{f:strange}
\end{figure}

There still remain a few test particles notable for their peculiar longer-timescale large inclination variations for which we have not identified any causative dynamical mechanism.  A typical example of these is shown in Figure~\ref{f:strange}: $i$ varies from $2^\circ$ to $9^{\circ}$ with a periodicity of about half-a-gigayear (with a corresponding synchronized variation in eccentricity from 0.06 to 0.15).  Test particle clones of four other CKBOs show similar orbital element time histories, but with shorter periodicities ranging from about 50 Myr to about 250 Myr.  These timescales are much longer than the libration timescales of MMRs which are typically $10^4$--$10^5$ yr \citep{Malhotra1996,Tiscareno2009}, or the timescale of the Kozai resonance within MMRs which is typically $10^6$--$10^7$ yr \citep{Chiang2003}.  The semimajor axis of the test particle is also inconsistent with the location of the $\nu_{8}, \nu_{18}$ and $\nu_{17}$ secular resonances which are all located at 40--42 AU \citep{Knezevic1991}.  We performed additional numerical simulations with the test particles exhibiting this peculiar behavior and were able to rule out numerical artifacts (we varied the integration timestep and integrated in both the ecliptic and invariable coordinate frames) as a potential cause.  We also tested for membership in all of the nearby resonances with Neptune, Uranus, and combinations of both planets that were identified by \citet{Nesvorny2001}, but found no librating resonance angles.  We speculate that the cause may be some other very high order MMR we did not consider, or possibly a super-resonance such as identified for Pluto by \citet{Williams1971} and \citet{Milani1989}; verification of this is left for future work.

\subsubsection{Resonant KBOs}\label{sss:resonant}

\begin{figure}
   \centering
   \includegraphics[width=5.5in]{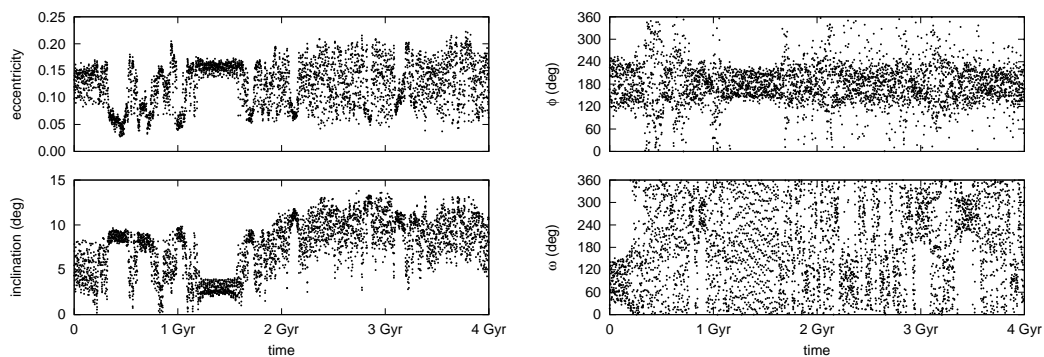} 
   \caption{Eccentricity, inclination, resonance angle ($\phi = 7\lambda_{kbo} - 4\lambda_{N} - 3\varpi_{kbo}$), and argument of perihelion, $\omega$, evolution over 4 Gyr for a test particle exhibiting intermittent libration within the 7:4 MMR with Neptune, as well as intermittent $\omega$ libration.}
   \label{f:kozai}
\end{figure}

The resonant objects in our simulation generally show more variation in $i$ than the non-resonant objects do;  large changes in $i$ ($\Delta i\ge 5^{\circ}$) occur for 51\% of the resonant objects over 4 Gyr compared to only 10\% of non-resonant objects (see Section~\ref{s:results}).  The variations in $i$ for the resonant KBOs occur on multiple timescales, with larger changes occurring on longer timescales;  this can be seen by comparing Figures~\ref{f:a_10Myrdi} and~\ref{f:a_di}.  Libration within an MMR causes variations in the orbital elements over timescales similar to the libration period of the resonance, which is typically a few tens of kiloyears for the resonances in this study.   Resonant particles also experience the same secular variations in $i$ on megayear-timescales that we describe above for the CKBOs.  Over longer timescales (a few tens of megayears), the Kozai resonance \citep{Kozai1962} accounts for many of the resonant particles having maximum inclination excursions $\Delta i\ge5^{\circ}$. The Kozai resonance is characterized by the libration of the argument of pericenter, and the preservation of the Kozai integral,
\begin{equation}
\Theta = \sqrt{(1-e^2)}\cos i.
\end{equation}
The effect of this resonance is large anti-correlated oscillations in eccentricity and inclination.  We find that many test particles do not librate in the Kozai resonance for the entire simulation, but instead experience intermittent libration which results in coupled large changes in $e$ and $i$.  We also find that the resonance angle $\phi$ (as defined by equation~\ref{eq:phi}) can exhibit intermittent changes in amplitude (correlated with intermittent Kozai libration) which also affects the inclination evolution.  A typical example of this type of evolution is shown in Figure~\ref{f:kozai}.  This type of intermittent libration of both $\phi$ and the argument of perihelion has been found in previous work exploring the dynamics of the 7:4 MMR in the classical belt region \citep{Lykawka2005r}.   This behavior is responsible for the stable resonant population having much larger maximum values of $\Delta i$ than the stable non-resonant CKBOs.

We also note that some members of the 7:4 and 9:5 MMRs diffuse out of resonance and evolve into stable orbits in the classical belt region. (These objects are indicated in Table~\ref{t:res_objects}).  This phenomenon likely contributes to the population of CKBOs that exhibit larger changes in $i$ (as discussed in Section~\ref{sss:ckbos}).  Objects like these `resonance escapees', whose dynamics at early times were dominated by the effects of MMRs, but later cease to librate in the resonance, are probably present in the observed sample of currently non-resonant objects.

\subsection{Reassessing the correlations between inclination and physical properties}\label{ss:correlations}

Here we examine the proposed correlations between physical properties and inclination using the results from our numerical simulations to represent the time-variability of the inclinations of the observed objects.  We use the Spearman rank correlation test \citep{Press1992} as a quantitative measure of the correlations between $i$ and each physical property.  This test has been used in previous studies of this problem (see \citet{Doressoundiram2008} and references therein); it is chosen because the results of the test do not depend on any assumed form for the correlation (i.e. the relationship need not be linear).  Values of $R_s$ close to $\pm1$ indicate a strong correlation or anti-correlation between the two parameters tested.  The significance level of the correlation is reported as the number of standard deviations above or below zero (no correlation) the value of $R_s$ lies compared to the null hypothesis that the two variables are uncorrelated.  For each physical property, we perform the correlation test using the average value of $i$ for the objects from our 10 Myr simulations, which is a better representation of their proper inclinations than the instantaneous observed value of $i$; we use the $rms$ $\Delta i$ from the 10 Myr simulation as a measure of the time variability of $i$ (using the average $i$ and $rms$ $\Delta i$ values from the 4 Gyr simulation does not alter the conclusions that we describe below).  The physical properties we consider for the correlation test are the spectral gradient and absolute magnitude; we do not consider albedo because only $\sim$ 30 CKBOs have measured albedos \citep{Brucker2009}.  We compare inclination distribution of the binary CKBOs to that of the apparently single KBOs, as well as the $i$ distribution of large CKBOs to the distribution determined by \citet{Brown2001} for the classical belt as a whole.  These comparisons are done using the Kolmogorov-Smirnov test (described in Section~\ref{s:modeling}) to measure the probability that the inclination distributions being compared are the same.

\subsubsection{Spectral Gradient}\label{sss:color}

\begin{figure}
   \centering
   \includegraphics{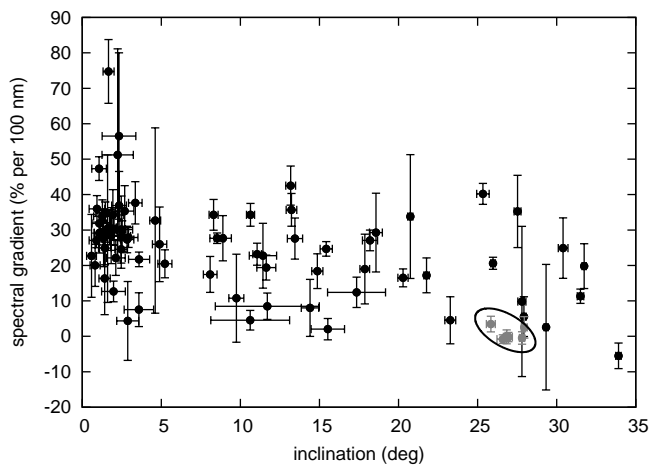} 
   \caption{Spectral gradient (percent reddening per 100 nm) vs.~average inclination over the 10 Myr simulation.  The error bars for the inclination are the $rms$ variation in $i$ over 10 Myr.  The 6 objects enclosed in the ellipse are all members of the Haumea collisional family.}
   \label{f:S_i}
\end{figure}

The colors of KBOs are assigned by photometric characterization of their reflectance spectra in various broadband filters (BVRI for the visible wavelengths).  The commonly used color indices (B-V, V-R, etc) represent the differences in the observed magnitude of the KBO in the two filters.  An additional color parameter that is determined from these broadband filter measurements is the spectral gradient, $S$; this is a measurement of the reddening of the KBO's reflectance spectrum over the wavelength range considered, usually expressed as percent reddening per 100 nm.  We use the spectral gradient in our color-$i$ correlation tests because $S$ can be calculated for any KBO that has been observed in at least two visible broadband filters, and it has been shown to correlate with the commonly measured V-R color for KBOs (see \citet{Doressoundiram2008} for a full discussion of this).  The results of our statistical tests do not depend on this choice (results for B-V, V-R, and B-R color indices are very similar to those for $S$), but using the spectral gradient provides us with the largest possible observational data set.  Figure~\ref{f:S_i} shows the spectral gradient vs.~10 Myr average inclinations for 80 CKBOs.  All values for $S$ are taken from the MBOSS database\footnote{http://www.eso.org/$\sim$ohainaut/MBOSS/} maintained by O.~Hainaut of the European Southern Observatory and described by \citet{Hainaut2002}.

In our calculations of the correlation coefficient, $R_s$, using the Spearman rank test, we want to include information about the observational uncertainties for the spectral gradient and the time variability of the inclinations (as represented by the $rms$ variations in $i$ calculated from the numerical simulations).  Including the uncertainties for both the physical property and the inclination is achieved by randomly assigning values to both the observed physical property and $i$, assuming that each variable has a normal distribution; this is repeated 1000 times and $R_s$ is calculated for each permutation of the data set.  The resulting range and significance of the values of $R_s$ yields a measure of how the observational errors in $S$ and the dynamical evolution of $i$ affect the correlation.  Using the average $i$ and its $rms$ value from the 10 Myr simulations as well as the observational errors in $S$, this procedure yields $R_s = -0.46 \pm 0.07$ ($4\pm0.5\sigma$ significance).  This can be compared to $R_s = -0.52$ ($4.6\sigma$ significance) for the test when we use only the best-estimate values of $S$ and the average values of $i$.  Most of the reduction of $R_s$ and its associated significance is due to the observational errors in $S$. When we account for the uncertainties in $S$ but do not take account of the time-variability of $i$, we find $R_s = -0.47 \pm 0.05$ ($4.2\pm0.5\sigma$ significance).  This is consistent with the fact that most of our observed sample of CKBOs with measured values of $S$ are stable and have fairly low variations in inclination. 

Another factor that could affect the correlation between spectral gradient and inclination is the existence of the Haumea (2003 EL$_{61}$) collisional family discovered by \citet{Brown2007}.  Haumea family members share a distinctive water ice absorption feature, and they are clustered in orbital element space with inclinations near $\sim25^{\circ}$;  the family members also tend to have neutral colors \citep{Schaller2008,Ragozzine2007}.  There are six Haumea family members (as defined by \citet{Brown2007} and \citet{Ragozzine2007}) in our set of CKBOs that have measured spectral gradients:  Haumea (2003 EL$_{61}$), 1995 SM$_{55}$, 1996 TO$_{66}$, 1999 OY$_{3}$, 2002 TX$_{300}$, and 2003 UZ$_{117}$.  These six objects comprise one third of the CKBOs with $i>25^{\circ}$ that have observed spectral gradients, and all six have neutral spectral gradients, creating a cluster in the plot of $S$ vs. $i$ (Figure~\ref{f:S_i}).  To test how this cluster of related objects affects the correlation coefficient for $S$ and $i$, we calculated $R_s$ for the CKBOs without these six objects in the data set.  Removing the Haumea family members, but not taking the uncertainty in $S$, or the time-variability of $i$ into account, we find $R_s = -0.42$ with a $3.6\sigma$ significance (compared to $R_s = -0.52$ with a $4.6\sigma$ significance with the Haumea family included).  If we remove the family members and account for the time-variability of $i$ and uncertainties in $S$, we find $R_s = -0.36 \pm 0.07$ ($3.1\pm0.6\sigma$ significance).  We conclude that the neutrally colored, high-inclination Haumea family is partially responsible for the reported correlation between color and inclination in previous studies.

The degree of correlation we find between color and inclination for the classical objects is smaller and less statistically significant than previously reported in the literature.  Several previous studies have found $R_s$ between $-0.6$ and $-0.7$ \citep{Jewitt2001,Doressoundiram2002,Trujillo2002,Peixinho2008};  the most recent of these \citep{Peixinho2008} analyzed the colors of 69 CKBOs, and reported $R_s=-0.7$ with a greater than $8\sigma$ significance.  In our analysis, even without accounting for the time-variability of $i$ or the observational errors in $S$, the inclusion of 11 newly observed CKBOs in the analysis in this work (most of which have large inclinations) lowers the value and significance of the correlation coefficient to $R_s = -0.52$ with $4.6\sigma$ significance.  This is a strong indication that the previously reported, $R_s = -0.7$ with $8\sigma$ significance, did not reflect the true correlation value and its significance for the intrinsic population;  if the significance level of the correlation were truly that high, it would have been extremely unlikely for additional observations to change the measured value of $R_s$ by such a large amount.  The large change in the significance of the correlation is likely a result of the observational incompleteness of the available sample of CKBOs, especially at higher inclinations.  Likewise, the inclusion or exclusion of the Haumea family members has a large effect on the significance of the correlation ($4 \sigma$ significance with the family, $3 \sigma$ without).  Therefore, while the trend of less spectral reddening with increasing inclination appears to be significant at the $3 \sigma$ level in the CKBO dataset, this result should be interpreted with caution until the addition of new observations does not drastically alter the results of the Spearman rank test.

We also tested the resonant KBOs for correlations between $S$ and $i$, but we found no statistically significant correlation ($R_s = -0.2 \pm 0.1$ significant at a $<1\sigma$ level), which is consistent with previous work on the color-$i$ relationship \citep{Doressoundiram2008}.  Unlike the case for the CKBOs, for the resonant objects the errors in $S$ contribute equally with the time-variability of $i$ to decrease both the value and significance of $R_s$ when compared to the use of only the best-estimate values of $S$ and $i$.  This is consistent with our numerical simulations, in which resonant objects undergo much larger changes in $i$ than CKBOs. Even if there were initial differences in spectral color between the low-$i$ and high-$i$ resonant populations, subsequent dynamical evolution would tend to erase the trend through the significant inclination variability of resonant KBOs that we found in our dynamical simulations.  

\subsubsection{Absolute Magnitude}\label{sss:size}

Here we examine the relationship between inclination and absolute magnitude. (In the absence of albedo measurements for most of the KBOs, their sizes remain uncertain, so absolute magnitude is used as a proxy for size).  The question of how these two properties might be related can be framed in several different ways: (1) Do the low-$i$ and high-$i$ populations have different size distributions? (2) Do the large and small KBOs have different inclination distributions? (3) What is the degree of correlation between the two parameters (the Spearman rank test)?  These are different questions, the answers to which have different implications.  \citet{Bernstein2004} addressed the first question and detected a difference between the size distributions of the low-$i$ and high-$i$ KBO populations; the implications this has for the accretional histories of these two populations has been explored in several works, most recently by \citet{Fraser2010}.  In the present work, we focus on the latter two questions; the differences in the inclination distributions of the population of large and small KBOs might tell us if the two groups experienced different dynamical histories. 

There are some indications that the largest objects in the classical belt have an inclination distribution that is different from that of the smaller objects;  \citet{Levison2001} report a statistically significant difference (97\% confidence level) between objects with absolute magnitudes $H\le6.5$ and those with $H>6.5$.  \citet{Levison2001} dismissed observational biases as the source of the difference in inclination distributions for large and small objects because the set of observed objects at that time did not show any correlations between size and ecliptic latitude at the time of discovery; discoveries of new KBOs at that time were dominated by ecliptic surveys.  This is no longer the case as off-ecliptic surveys now account for many KBO discoveries, especially at brighter magnitudes.  The observational surveys most capable of detecting faint objects have been performed near the ecliptic where there is a strong preference for finding low-inclination objects;  the wider latitude surveys have brighter limiting magnitudes, and are therefore not as likely to detect small objects.  The result is that observations of the brightest objects have fairly evenly sampled the inclination distribution of the brighter objects, while the observations of the fainter objects are much more heavily sampled at low inclinations.  Any test we use to compare the two distributions (one complete, and one biased) will show a difference between the distributions because of this sampling bias.  A better way to test whether the largest objects have the same inclination distribution as the smaller objects is to compare the $i$ distribution of the largest CKBOs to a debiased $i$ distribution for all CKBOs.  Here we will make the comparison using both the best-fit model to our debiased inclination distribution (the initial distribution shown in Figure~\ref{f:itime}) and the best fit model distribution calculated by \citet{Brown2001} for the classical belt.

To minimize the effect of observational biases on the correlation analysis, we define the group of large CKBOs by adopting an absolute magnitude cutoff that yields a set of objects that is observationally fairly complete.  Wide area surveys have detected most objects with apparent magnitudes $R < 21$ within $30^{\circ}$ of the ecliptic \citep{Trujillo2003,Schwamb2009}.  The apparent magnitude $R\approx 21$ corresponds to an absolute magnitude of $H\approx5$ at 40 AU, so we use this as the cutoff for observational completeness.  At this absolute magnitude there is a sharp drop off in the fraction of objects detected at large ecliptic latitudes:  60\% of objects with $H<5$ were detected at ecliptic latitudes between 5 and 30$^{\circ}$ while only 15\% of objects with $5<H<6$ found at similar latitudes.  \citet{Levison2001} reported that CKBO's with $H<6.5$ tend to have higher inclinations; if that trend is real, then the conclusions should hold for our smaller $H$ cutoff (larger minimum size).  There are 26 CKBOs with $H\le5$ in our CKBO sample (even with a more strict size cutoff, this is a significant improvement over the 8 objects with $H\le6.5$ available to \citet{Levison2001} for their analysis). In Section 3, we noted that a natural cutoff inclination between the low and high inclination populations is $i=5^\circ$ for our best-fit double gaussian model, and $i=7^\circ$ for Brown's 2001 model.  Of the 26 large CKBOS, only one has $i\le5^{\circ}$ and three have $i\le7^{\circ}$.      We can compare these observed numbers of objects to the expected number of low-$i$ CKBOs.  Using our best fit parameters  and their $1\sigma$ uncertainties, we expect between 4 and 8 objects with $i\le5^{\circ}$ in a set of 26 observations.  The probability of observing only one such object is $4^{+2}_{-3.6}\%$.  Using the \citet{Brown2001} best fit parameters and $1\sigma$ uncertainties, we would expect between 5 and 8 CKBOs with $i\le7^{\circ}$ compared to the 3 we observe.  The probability of this happening by chance is $7^{+12}_{-4}\%$.  As noted in Section~\ref{s:results}, neither of these two model fits is a very good description of our debiased inclination distribution, but they do offer some description of the expected ratios of high- to low-inclination CKBOs.  The subset of CKBOs with $H\le5$ contains fewer low-$i$ objects than expected if the true inclination distribution for this brightness range contains both high- and low-$i$ Gaussian components; from the above analysis, we can estimate that there is a $4-7\%$ probability ($0.4-19\%$ when we account for the uncertainties in the model fits) that this observation arises by chance.  Considering our full range of uncertainties, this result is consistent with the 3\% reported by \citet{Levison2001}.

If the inclination distribution varies with size, we also might expect to see a correlation between $H$ and $i$.  For the smaller CKBOs, with $H>5$, the correlation coefficient between $H$ and average $i$ from the 10 Myr simulations we find is $R_s=-0.07$ ($1.3 \sigma$ significance); when we include the effect of inclination variability from the 10 Myr simulations (as described in the previous section), we find $R_s=0.07\pm0.01$ ($1.3\pm0.3 \sigma$ significance).  This result is consistent with no correlation between $H$ and $i$ for $H>5$.   For the entire set of CKBOs, the correlation coefficient drops to $R_s = 0.04\pm0.09$, which is also consistent with no correlation.

For the larger CKBOs, with $H\le5$, the correlation coefficient between $H$ and average $i$ from the 10 Myr simulations is $R_s=-0.3$ ($1.5 \sigma$ significance); including the time variability of $i$, we find $R_s=-0.29\pm0.02$ ($1.5\pm0.1 \sigma$ significance).  These relatively low values of $|R_s|$, together with the low $\sim1.5\sigma$ significance level of the Spearman rank correlation means that we do not have a definitive answer to whether $H$ and $i$ are correlated such that there is a different inclination distribution at different sizes of the KBOs.  The anti-correlation between $H$ and $i$ is not highly significant for the large objects.   There are fewer large objects than expected at small $i$, but there is a non-negligible $\sim4-7\%$ probability that the discrepancy is a result of the relatively small number of $H\le5$ objects rather than a true difference in the inclination distributions.   Future wide-area surveys with deeper limiting magnitudes will allow us to compare the $i$ distribution for a larger observationally complete set of bright objects to the classical belt inclination distribution, at which time a more definitive judgement can be made.

For the resonant KBOs, we find no significant correlation between $H$ and $i$ for the 194 observed objects.  Employing the same method for including the time-variability of $i$ into the calculation of the correlation coefficient as in Section~\ref{sss:color}, we find $R_s = (-2\pm2)\times10^{-2}$, consistent with the two parameters being unrelated.

\subsubsection{Fraction of Binary Objects}\label{sss:binarity}

\citet{Noll2008} searched 101 known CKBOs with inclinations ranging from $0.6-34^{\circ}$ for binary companions and found a binary fraction of $\sim30\%$ for objects at $i<5.5^{\circ}$ compared to a binary fraction of $\sim10\%$ at higher inclinations;  if the comparison is restricted to the fraction of objects that are same-size binaries, the binary fractions are $\sim30\%$ and $\sim2\%$ respectively.  They used the Kolmogorov-Smirnov test to evaluate the significance of the differences between the $i$-distributions of binary and (apparently) non-binary objects, finding a $4.7\%$ probability that the two populations were drawn from the same distribution.  Here we redo the Kolmogorov-Smirnov test, taking into account the time-variability of the inclinations.  Of the 101 objects in \cite{Noll2008}, 93 are in our sample of CKBOs; 21 of these are binary.  We run the K-S test 1000 times, each time randomly assigning values of the inclinations from a normal distribution having average and standard deviation in $i$ found in our 10 Myr simulation.  This calculation finds that there is a $6.5\pm3\%$ probability that the binary and non-binary CKBOs are derived from the same overall inclination distribution.  This result is consistent with \citet{Noll2008}'s results. We conclude that the time-variability of inclinations in the classical Kuiper belt does not have a strong effect on the inclination dependency of the binary fraction.

\section{Summary}\label{s:summary}

We studied the dynamics of the presently observed classical and resonant Kuiper belt objects, with a focus on their inclination evolution.

\begin{enumerate}

\item Planetary perturbations over gigayear timescales can produce inclination changes in the orbits of Kuiper belt objects.  We find that large changes ($\Delta i \ge 5^{\circ}$) are confined to a small fraction of the parameter space, and they typically last only a relatively small fraction of the time over $\sim$~4 gigayears.  Most objects (90\% of CKBOs and 77\% of resonant KBOs) remain within $5^{\circ}$ of their current inclinations and have short term variations with amplitude typically $<2^{\circ}$ (relative to the invariable plane).  Overall, the debiased inclination distribution of the non-resonant classical Kuiper belt is nearly stationary under planetary perturbations over gigayear timescales. 

\item Considering the inclination distribution of classical Kuiper belt objects, we find that there is some transfer between the high- and low-$i$ populations, but at any given time relatively few CKBOs ($<5\%$) will be found at inclinations inconsistent with their original inclination classification.  Most CKBOs that experience changes in $i$ larger than $5^{\circ}$ are ejected from the classical belt by close encounters with Neptune, but some objects (mostly in the outer half of the classical belt) can undergo large inclination changes while maintaining fairly low eccentricities thus avoiding close Neptune encounters.

 \item  We identify many KBOs  within the classical belt region that are in high-order MMRs with Neptune.  We find that some current members of the 7:4 and 9:5 MMRs escape from these resonances into the non-resonant classical belt population without having destabilizing close encounters with Neptune.  Many of the CKBOs in our simulation that experience large changes in $i$ are located near (but not in) these high-order resonances.  Temporary capture into high-order MMRs could explain the CKBOs with the most changeable inclinations.

 \item The resonant KBOs are more likely to undergo large changes ($\Delta i > 5^{\circ}$) in inclination than non-resonant objects.  The largest variations in $i$ can be attributed to the Kozai mechanism operating in conjunction with a mean motion resonance.  We estimate that, at any given epoch, $5$ to $10\%$ of observed resonant objects will be at inclinations inconsistent with their original inclination classification.

\item We reassessed the previously proposed correlations between inclination and the binarity, sizes, and colors of classical Kuiper belt objects, using updated observations and taking account of the time-variability of the inclinations.  

\begin{enumerate}

\item The inclination time-variability does not greatly affect \citet{Noll2008}'s finding that the binary and non-binary CKBOs have different inclination distributions; we find only a $6.5\pm3\%$ probability that the binary and non-binary CKBOs are derived from the same inclination distribution, very similar to the $\sim5\%$ probability found by \citet{Noll2008}.    

\item For the observationally near-complete population of bright objects (absolute magnitudes $H\le5$), we find a $4-7\%$ probability that the large objects have the same inclination distribution as the rest of the CKBOs.  This is a slightly higher probability than the previously reported $3\%$ chance of similarity \citep{Levison2001}, although the two results are consistent within our error estimates for the probability; most of the difference is due to the larger data set available since the previous study.  Using the Spearman rank test, we find no significant correlation between $i$ and $H$ for the CKBOs as a whole or for the bright ($H\le5$) CKBOs considered separately.

\item We find a decreased significance of the correlation between color (represented here by spectral gradient) and inclination of CKBOs;  previous estimates reported the correlation coefficient $R_s = -(0.6-0.7)$ ($8\sigma$ significance) while our analysis finds $R_s = -0.46 \pm 0.07$ ($4\pm0.5\sigma$ significance).  A large part of the difference between our result and the previous results is due to the larger data set of observed objects, with a small (but significant) contribution from taking account of the observational uncertainty in the measured colors and the time-variability of $i$.  There is also evidence that the Haumea collisional family is responsible for at least part of the apparent correlation;  removing this high-inclination, neutrally colored group of objects lowers the correlation and its significance to $R_s = -0.36\pm0.07$ ($3.1\pm0.6\sigma$ significance). 
\end{enumerate}

\end{enumerate}

\acknowledgments
This research was supported by grant no. NNX08AQ65G from NASA's Outer Planets Research program.  

\clearpage

\bibliographystyle{apj}
\bibliography{ms}

\begin{deluxetable}{l l l l l l}
\tabletypesize{\footnotesize}
\tablecolumns{6}
\tablewidth{442pt}
\tablecaption{Resonant Objects}
\tablehead{ \colhead{MMR} & \multicolumn{5}{c}{Designations of Members} }

\startdata
3:2
& 1993 RO &
1993 SB &
1993 SC &
1994 JR$_{1}$ &
1994 TB \\
& 1995 HM$_{5}$ &
1995 QY$_{9} (r)$ &
1995 QZ$_{9}$ &
1996 RR$_{20}$ &
1996 SZ$_{4}$ \\ 
& 1996 TP$_{66}$ &
1996 TQ$_{66}$ &
1997 QJ$_{4}$ &
1998 HH$_{151}$ &
1998 HK$_{151}$ \\
& 1998 HQ$_{151}$ &
1998 UR$_{43}$ &
1998 US$_{43}$ (r) &
1998 VG$_{44}$ &
1998 WS$_{31}$ \\
& 1998 WU$_{31}$ & 
1998 WV$_{31}$  &
1998 WW$_{24}$ &
1998 WZ$_{31}$ &
1999 CE$_{119}$ \\ 
& 1999 CM$_{158}$ (r) &
1999 RK$_{215}$ &
1999 TC$_{36}$ &
1999 TR$_{11}$ &
2000 CK$_{105}$ \\
& 2000 EB$_{173}$ &
2000 FB$_{8}$ &
2000 FV$_{53}$ (r)&
2000 GE$_{147}$ &
2000 GN$_{171}$ \\
& 2000 YH$_{2}$ &
2001 FL$_{194}$ &
2001 FR$_{185}$ &
2001 FU$_{172}$ &
2001 KB$_{77}$ \\
& 2001 KD$_{77}$ &
2001 KN$_{77}$ (r)&
2001 KQ$_{77}$ &
2001 KX$_{76}$ &
2001 KY$_{76}$ \\
& 2001 QF$_{298}$ & 
2001 QG$_{298}$ &
2001 QH$_{298}$ &
2001 RU$_{143}$ &
2001 RX$_{143}$ \\
& 2001 UO$_{18}$ (r) &
2001 VN$_{71}$ &
2001 YJ$_{140}$ &
2002 CE$_{251}$ (r)&
2002 CW$_{224}$ \\
& 2002 GE$_{32}$ &
2002 GF$_{32}$ &
2002 GL$_{32}$ &
2002 GV$_{32}$ &
2002 GW$_{31}$ \\
& 2002 GY$_{32}$ &
2002 VD$_{138}$ & 
2002 VE$_{95}$ & 
2002 VR$_{128}$ &
2002 VU$_{130}$ \\
& 2002 VX$_{130}$ &
2002 XV$_{93}$ &
2003 AZ$_{84}$ &
2003 FB$_{128}$ &
2003 FF$_{128}$ \\
& 2003 FL$_{127}$ &
2003 HA$_{57}$ &
2003 HD$_{57}$ &
2003 HF$_{57}$ &
2003 QB$_{91}$ \\ 
& 2003 QH$_{91}$ &
2003 QX$_{111}$ &
2003 SO$_{317}$ &
2003 SR$_{317}$ &
2003 TH$_{58}$ (r)\\
& 2003 UT$_{292}$ &
2003 UV$_{292}$ &
\textbf{2003 UZ$_{413}$} &
2003 VS$_{2}$ &
2003 WA$_{191}$ \\
& 2003 WU$_{172}$ (r)&
2004 DW &
2004 EH$_{96}$ &
2004 EJ$_{96}$ &
\textbf{2004 EV$_{95}$} \\
& 2004 EW$_{95}$ (r) &
2004 FU$_{148}$ &
2004 FW$_{164}$ &
\textbf{2004 US$_{10}$} &
\textbf{2004 VT$_{75}$} \\
& \textbf{2004 VZ$_{75}$} &
2005 EZ$_{296}$ &
2005 EZ$_{300}$ (r) &
2005 GA$_{187}$ &
2005 GB$_{187}$ \\
& 2005 GE$_{187}$ &
2005 GF$_{187}$ &
\textbf{2005 GV$_{210}$} &
2005 TV$_{189}$ &
\textbf{2006 HJ$_{123}$} \\
& \textbf{2007 JF$_{43}$} &
\textbf{2007 JH$_{43}$} (r)& 
Pluto & & \\ 

7:4 
& \textbf{(1994 GV$_{9}$)} &
 1999 CD$_{158}$ (r) &
 \textbf{1999 CO$_{153}$}  &
 1999 HG$_{12}$ &
(1999 HR$_{11}$) \\
& 1999 HT$_{11}$  &
 1999 KR$_{18}$ &
 1999 RH$_{215}$ &
 2000 FD$_{8}$ (r) &
 (2000 FX$_{53}$) \\
& 2000 OP$_{67}$  &
 2000 OY$_{51}$ (r) &
 2000 YU$_{1}$ (r) &
 2001 HA$_{59}$ (r) & 
 2001 KJ$_{76}$ \\
& 2001 KO$_{76}$ &
2001 KP$_{76}$ (r) &
 2001 KP$_{77}$ (r) &
 2001 QE$_{298}$ &
 2002 PA$_{149}$ (r) \\
& 2002 PB$_{171}$ & 
 2003 QW$_{111}$ &
 2003 QX$_{91}$ &
 \textbf{(2003 YJ$_{179}$)}  &
  2004 OK$_{14}$ (r) \\
 & 2004 PW$_{107}$ (r) &
 \textbf{2004 VU$_{75}$} (r) &
 \textbf{2005 SF$_{278}$} (r) & & \\

2:1
& 1996TR$_{66}$ (r)&
1997SZ$_{10}$ (r)&
1998SM$_{165}$ (r) &
1999RB$_{215}$ (r) &
1999RB$_{216}$ (r)\\
& 2000JG$_{81}$ (r)&
2000QL$_{251}$ &
2001FQ$_{185}$ &
2001UP$_{18}$ &
2002PU$_{170}$ (r) \\
& 2002WC$_{19}$ &
2002VD$_{130}$ &
2003FE$_{128}$ (r) &
2004TV$_{357}$ & 
\textbf{2005CA$_{79}$} \\
& \textbf{2005RS$_{43}$} & & & &\\

5:3
& 1994 JS (r) &
 {\textbf{1997 CV$_{29}$}} &
 1999 CX$_{131}$ (r) &
 2000 PL$_{30}$ (r) &
 2000 QN$_{251}$ (r) \\
& 2001 XP$_{254}$ (r) &
 2001 YH$_{140}$ (r) &
 2002 GS$_{32}$ &
2002 VA$_{131}$ &
 2002 VV$_{130}$ \\
& 2003 US$_{292}$ (r) &
 2003 YW$_{179}$ &
 {\textbf{2005 SE$_{278}$}} &
 {\textbf{2005 TN$_{74}$}} (r) &
 \textbf{2006 QQ$_{180}$}   \\

9:5
& \textbf{1999 CS$_{153}$} (r) &
 \textbf{2000 CN$_{105}$} (r) &
 \textbf{(2000 FR$_{53}$)} &
 \textbf{2000 YZ$_{1}$} (r) &
 2001 KL$_{76}$ \\
& 2002 GD$_{32}$ &
 \textbf{(2004 VS$_{75}$)} &
 \textbf{(2005 JH$_{177}$)} & & \\ 
 
4:3 
& 1995 DA$_{2}$ &
 1998 UU$_{43}$ &
 1999 RW$_{215}$ &
  2002 FW$_{6}$ &
   2003 SS$_{317}$ (r) \\*
  & \textbf{2004 TX$_{357}$} (r) &
  \textbf{2005 ER$_{318}$} (r) & & &\\ 
 
5:4 
& 1999 CP$_{133}$ & 
 2001 XH$_{255}$  & 
 2002 GW$_{32}$ (r) &
 2003 FC$_{128}$ &
 2003 QB$_{92}$ \\
& \textbf{2005 SC$_{278}$} & & & &\\ 

11:6 &  
\textbf{1999 CP$_{153}$} (r) &
2001 KU$_{76}$ &
\textbf{2005 EB$_{318}$} (r) & & \\ 
 
8:5 &
\textbf{2005 VZ$_{122}$} (r) & & & &\\ 

11:8 &
2001 XS$_{254}$ (r) & & & &\\ 

16:9 &
\textbf{2005 GC$_{187}$} (r) & && & \\ 
\enddata

\tablecomments{Entries in boldface indicate objects not previously identified as resonant by \citet{Gladman2008}, \citet{Lykawka2007}, or \citet{Chiang2003}.  Entries in parentheses are objects whose test particles exit the resonance by the end of the 4Gyr simulation without being removed from the classical belt region.  Entries followed by (r) indicate objects whose test particles are removed from the simulation after close encounters with Neptune.  }
\label{t:res_objects}
\end{deluxetable}

\begin{deluxetable}{c l l}
\tabletypesize{\footnotesize}
\tablecolumns{3}
\tablewidth{0pt}
\tablecaption{Kuiper Belt Surveys}
\tablehead{ \colhead{Number} & \colhead{Survey} & \colhead{Reference(s)}\\
\colhead{of Objects} & &  }

\startdata
171 		& Deep Ecliptic Survey 					& \citet{Elliot2005} \\ 
34 		& Canada-France Ecliptic Plane Survey 		& \citet{Jones2006} \\
 		&									& \citet{Kavelaars2009} \\
33 		& Canada-France-Hawaii Telescope Survey 	& \citet{Trujillo_CFHT2001} \\
8 		& Spacewatch 							& \citet{Larsen2001} \\
 		& 									& \citet{Larsen2007} \\ 
7 		& Mauna Kea 							& \citet{Jewitt1998} \\ 
6 		& Distant Disk Survey 					& \citet{Allen2002} \\ 
19 		& miscellaneous						& \citet{Irwin1995} \\
		&									& \citet{Jewitt1995} \\
		&									&  \citet{Jewitt1996} \\
		&									& \citet{Gladman1998} \\
		&									& \citet{Luu1998} \\
		&									& \citet{Trujillo1998} \\
		&									& \citet{Gladman2001} \\
		&									& \citet{Trujillo2001} \\ 
 		& 									& \citet{Trujillo2003} \\ 
\enddata

\tablecomments{Surveys used in this study to construct a debiased model of the non-resonant classical Kuiper belt.}

\label{t:surveys}
\end{deluxetable}

\end{document}